# The diffusion of charged particles in the weakly ionized plasma with power-law kappa-distributions

Lan Wang and Jiulin Du*

*Department of Physics, School of Science, Tianjin University, Tianjin* 300072, *China*

We study the diffusion of charged particles in the weakly ionized plasma with the power-law $\kappa$-distributions and without magnetic field. The electrons and ions have different $\kappa$-parameters. We obtain the expressions of both diffusion and mobility coefficients of electrons and ions respectively in the plasma. We find that these new transport coefficient formulae depend strongly on the $\kappa$-parameters in the power-law distributed plasma. When we take $\kappa\to\infty$, these formulae reduce to the classical forms in the weakly ionized plasma with a Maxwellian distribution.

## I. Introduction

The $\kappa$-distribution is a power-law distribution function with a parameter $\kappa$, which was first drawn by Vasyliunas in 1968 to model the velocity distribution of the electrons at high energies in the energy spectra of observed electrons within the plasma sheet in the magnetosphere.[1] It can be expressed by

$$f_\kappa(\varepsilon) = n_0 B_k \left[1 + \frac{1}{\kappa}\frac{\varepsilon}{\varepsilon_0}\right]^{-(\kappa+1)}, \tag{1}$$

where $\varepsilon = \frac{1}{2}mv^2$ is kinetic energy of an electron with speed $v$ and mass $m$, $\varepsilon_0 = \frac{1}{2}mw_\kappa^2$ is a characteristic energy, with a characteristic speed $w_\kappa$ related to the most probable speed $w_0$ of a Maxwellian distribution with equal number and energy densities by $w_\kappa = w_0[(\kappa-\frac{3}{2})/\kappa]^{1/2}$,[2] and $B_\kappa$ is the normalization constant,

$$B_\kappa = \left[\pi w_0^2(\kappa - \frac{3}{2})\right]^{-3/2} \frac{\Gamma(\kappa+1)}{\Gamma(\kappa-\frac{1}{2})}, \tag{2}$$

where $w_0 \equiv (2k_B T/m)^{1/2}$ with temperature $T$ and Boltzmann constant $k_B$. Obviously, the parameter $\kappa > \frac{3}{2}$ and in the limit $\kappa\to\infty$ the $\kappa$-distribution (1) reduces to a Maxwellian distribution.

For convenience, the $\kappa$-distribution can also be written as

$$f_\kappa(v) = n_0 B_\kappa \left(1 + A_\kappa v^2\right)^{-(\kappa+1)} \tag{3}$$

with

$$A_\kappa = \frac{m}{(2\kappa-3)k_B T}.$$

Non-Maxwellian distributions can be observed commonly both in space plasmas and in laboratory plasmas. For instance, the plasmas in the planetary magnetospheres are found to be deviation from Maxwellian distribution due to the presence of high energy particles.[2] The spacecraft measurements of plasma velocity distributions, both in the solar wind and in the planetary magnetospheres and magnetosheaths, have revealed that non-Maxwellian distributions are quite common. In many situations the distribution have a "suprathermal" power-law tail at high energies, which has been well modeled by the $\kappa$-distribution.[3]

* Email: jldu@tju.edu.cn

In recent years, the researches on the plasmas with the $\kappa$-distribution as well as the $\kappa$-like power-law distributions have attracted great interest for their many interesting applications found in the wide fields of space plasma physics and astrophysics, and also for the $\kappa$-distribution family that can be studied under the framework of nonextensive statistics.[4-12] And with the aid of the nonextensive kinetic theory, one can determine the expression of the $\kappa$-parameter and its physical meaning in the astrophysical and space plasmas.[10]

The transport coefficients in the $\kappa$-distributed plasma were first studied under a very simplified Lorentz model,[7] containing electric conductivity, thermoelectric coefficient and thermal conductivity. In this work, we will study the diffusion process of the weakly ionized plasma with the $\kappa$-distribution and calculate the diffusion coefficients. The weakly ionized plasma contains less charged particle than neutral particle, and the interactions between particles are taking place by the collisions, including electron-electron collision, electron-ion collision, electron-neutral particle collision and ion-neutral particle collision. Since the number of charged particles is small, when we calculate the transport coefficients we mainly consider the collisions of charged particles with the neutral particles and neglect the collisions between the charged particles.[13]

This paper is organized as follows. In sec.II, we introduce the Boltzmann transport equation for the weakly ionized plasma with $\kappa$-distributions. In sec.III, we derive the expressions of diffusion coefficients of charged particles in the plasma with the $\kappa$-distributions. In sec.IV, we derive the mobility coefficient of charged particles in the plasma with the $\kappa$-distributions. Finally in sec.V, we give conclusion.

## II. Transport equation

The problem of plasma transport is often studied by a model rather than by a rigorous treatment. Such a model can not only simulate the physical characteristics of the two-body interactions, but also provide some mathematical simplifications. For weakly ionized plasmas, the model postulates that the deflection of a charged particle is caused by a single collision with a neutral atom rather than multiple scattering of other charged particles. In this case, the Boltzmann form of the collision integral is correct, and this model together with further simplifications is used to find the transport properties of weakly ionized plasmas. And the near collisions between the ionized plasma particles and the neutral background gas are the main mechanism for determining diffusion, electric conductivity, and so on.

In weakly ionized plasma (without magnetic field), a collision between a plasma particle and a neutral molecule or atom can generally be treated as an elastic collision.

Moreover, the duration of such a collision is generally much smaller than the time interval between the two collisions. This fact, together with the molecular chaos hypothesis constitutes the basis for a statistical model of low-density gas, and the mathematical representation of this model is the Boltzmann transport equation. So the statistical description of this multi-body system has the following form:[14]

$$\frac{\partial f_\alpha}{\partial t} + \mathbf{v}_\alpha \cdot \frac{\partial f_\alpha}{\partial \mathbf{r}} + \frac{e\mathbf{E}}{m_\alpha} \cdot \frac{\partial f_\alpha}{\partial \mathbf{v}_\alpha} = \left( \frac{\partial f_\alpha}{\partial t} \right)_C . \tag{4}$$

This is the Boltzmann equation, where $f_\alpha(\mathbf{r},\mathbf{v},t)$ is a single-particle velocity distribution function at time $t$, velocity $\mathbf{v}$ and position $\mathbf{r}$, the subscript $\alpha = e$, $i$ is electron and ion respectively, and $\mathbf{E}$ is the electric field. The term on the right-hand side is the collision term, which represents the change of the distribution function due to the collisions between particles.

In weakly ionized plasma, because the neutral particles are heavy as compared with electrons and ions, they can be regarded as static and homogeneous distribution, and we mainly consider the collisions between charged particles and neutral particles and neglect the collision between the charged particles. In this way, here we can employ the Krook collision model.[14] That is

$$\begin{aligned}\left( \frac{\partial f_\alpha}{\partial t} \right)_C &= -\frac{1}{\tau_\alpha}\left( f_\alpha - \frac{n_\alpha}{n_{\alpha 0}} f_\alpha^{(0)} \right) \\ &= -\nu_\alpha \left( f_\alpha - \frac{n_\alpha}{n_{\alpha 0}} f_\alpha^{(0)} \right),\end{aligned} \tag{5}$$

where $\tau_\alpha$ is a mean time of collision between the charged particles and the neutral particles. Or if $\nu_\alpha$ is the mean collision frequency, we have that $\tau_\alpha=(\nu_\alpha)^{-1}$. $n_{\alpha 0} = \int f_\alpha^{(0)} d\mathbf{v}_\alpha$ is the particle density in a equilibrium state and $n_\alpha = \int f_\alpha d\mathbf{v}_\alpha$ is the local particle density in a nonequilibrium state.

Usually we can write the velocity distribution function as the following form,

$$f_\alpha = \frac{n_\alpha}{n_{\alpha 0}} f_\alpha^{(0)} + f_\alpha^{(1)}, \tag{6}$$

where $f_\alpha^{(1)}$ is a small disturbance about the stationary state distribution $f_\alpha^{(0)}$. In this function we replace the distribution function with uniform particle density by the distribution function with local particle density to guarantee the conservation of particle number. And the stationary state distribution is the $\kappa$-distribution,

$$f_\alpha^{(0)} = f_{\kappa,\alpha} = n_{\alpha 0} B_{\kappa,\alpha} \left( 1 + A_{\kappa,\alpha} {v_\alpha}^2 \right)^{-(\kappa_\alpha + 1)}, \tag{7}$$

with

$$B_{\kappa,\alpha} = \left[ \frac{2\pi k_B T_\alpha}{m_\alpha} (\kappa_\alpha - \frac{3}{2}) \right]^{-3/2} \frac{\Gamma(\kappa_\alpha + 1)}{\Gamma(\kappa_\alpha - \frac{1}{2})},$$

and

$$A_{\kappa,\alpha}=\frac{m_\alpha}{\left(2\kappa_\alpha-3\right)k_BT_\alpha}. \tag{8}$$

Submitting Eq.(5) and Eq.(6) into Eq.(4), we can obtain that

$$\frac{\partial f_{\kappa,\alpha}}{\partial t}+\mathbf{v}_\alpha\cdot\frac{\partial f_{\kappa,\alpha}}{\partial \mathbf{r}}+\mathbf{a}_\alpha\cdot\frac{\partial f_{\kappa,\alpha}}{\partial \mathbf{v}_\alpha}=-\nu_\alpha f_{\kappa,\alpha}^{(1)}. \tag{9}$$

Since $f_{\kappa,\alpha}^{(1)}<<\frac{n_\alpha}{n_{\alpha 0}}f_{\kappa,\alpha}^{(0)}$, we can neglect $\partial f_{\kappa,\alpha}^{(1)}/\partial r$ and $\partial f_{\kappa,\alpha}^{(1)}/\partial v_\alpha$, then the equation becomes

$$f_{\kappa,\alpha}^{(1)}=-\frac{1}{\nu_\alpha}\left[\mathbf{v}_\alpha\cdot\frac{\partial}{\partial\mathbf{r}}\left(\frac{n_\alpha}{n_{\alpha0}}f_{\kappa,\alpha}\right)+\mathbf{a}_\alpha\cdot\frac{\partial}{\partial\mathbf{v}_\alpha}\left(\frac{n_\alpha}{n_{\alpha0}}f_{\kappa,\alpha}\right)\right], \quad 10)$$

and so we have that

$$f_\alpha=\frac{n_\alpha}{n_{\alpha0}}f_{\kappa,\alpha}-\frac{1}{\nu_\alpha}\left[\mathbf{v}_\alpha\cdot\frac{\partial}{\partial\mathbf{r}}\left(\frac{n_\alpha}{n_{\alpha0}}f_{\kappa,\alpha}\right)+\mathbf{a}_\alpha\cdot\frac{\partial}{\partial\mathbf{v}_\alpha}\left(\frac{n_\alpha}{n_{\alpha0}}f_{\kappa,\alpha}\right)\right], \tag{11}$$

where we have used $\mathbf{a}_\alpha=e\mathbf{E}/m_\alpha$ .

**III. The diffusion coefficient of charged particles**

Transport processes in nonequilibrium plasma involve a variety of thermodynamic “fluxes”, such as heat flow, electric current and particle flow etc. These “fluxes” are driven by the corresponding nonequilibrium thermodynamic gradients; e.g. heat flow is driven by temperature gradient, electric current is driven by electric potential gradient, and particle flow is driven by density gradient. The particle flow density vector is expressed by the distribution function as[15]

$$\begin{aligned}\mathbf{\Gamma}_\alpha&=n_\alpha\left\langle\mathbf{v}_\alpha\right\rangle=\int\mathbf{v}_\alpha f_\alpha d\mathbf{v}_\alpha\\&=\frac{n_\alpha}{n_{\alpha0}}\int\mathbf{v}_\alpha f_{\kappa,\alpha}d\mathbf{v}_\alpha-\int\mathbf{v}_\alpha\left[\frac{\mathbf{v}_\alpha}{\nu_\alpha}\cdot\frac{\partial}{\partial\mathbf{r}}\left(\frac{n_\alpha}{n_{\alpha0}}f_{\kappa,\alpha}\right)+\frac{\mathbf{a}_\alpha}{\nu_\alpha}\cdot\frac{\partial}{\partial\mathbf{v}_\alpha}\left(\frac{n_\alpha}{n_{\alpha0}}f_{\kappa,\alpha}\right)\right]d\mathbf{v}_\alpha .\end{aligned} \tag{12}$$

Because $f_{\kappa,\alpha}$ is an even function about the speed **v**, the first integral on the right-hand side of Eq.(11) is equal to zero. Then the particle flow density vector becomes

$$\mathbf{\Gamma}_\alpha=-\int\mathbf{v}_\alpha\left[\frac{\mathbf{v}_\alpha}{\nu_\alpha}\cdot\frac{\partial}{\partial\mathbf{r}}\left(\frac{n_\alpha}{n_{\alpha0}}f_{\kappa,\alpha}\right)+\frac{\mathbf{a}_\alpha}{\nu_\alpha}\cdot\frac{\partial}{\partial\mathbf{v}_\alpha}\left(\frac{n_\alpha}{n_{\alpha0}}f_{\kappa,\alpha}\right)\right]d\mathbf{v}_\alpha, \tag{13}$$

where the first term is the diffusion flow due to thermal motion, and the second term is due to electric field driven particle migration flow.

If the diffusion flow is denoted by $\Gamma_{\alpha,D}$ , then

$$\begin{aligned}\mathbf{\Gamma}_{\alpha,D}&=-\int\mathbf{v}_\alpha\left[\frac{\mathbf{v}_\alpha}{\nu_\alpha}\cdot\frac{\partial}{\partial\mathbf{r}}\left(\frac{n_\alpha}{n_{\alpha0}}f_{\kappa,\alpha}\right)\right]d\mathbf{v}_\alpha\\&=-\frac{1}{n_{\alpha0}\nu_\alpha}\int\mathbf{v}_\alpha f_{\kappa,\alpha}\mathbf{v}_\alpha\cdot\frac{\partial n_\alpha}{\partial\mathbf{r}}d\mathbf{v}_\alpha\end{aligned} \tag{14}$$

If we let the velocity be that

$$\mathbf{v}_\alpha = v_x \boldsymbol{i} + v_y \boldsymbol{j} + v_z \boldsymbol{k}, \tag{15}$$

and

$$\frac{\partial n_\alpha}{\partial \mathbf{r}} = \frac{x}{r}\frac{\partial n_\alpha}{\partial r}\boldsymbol{i} + \frac{y}{r}\frac{\partial n_\alpha}{\partial r}\boldsymbol{j} + \frac{z}{r}\frac{\partial n_\alpha}{\partial r}\boldsymbol{k}, \tag{16}$$

we therefore have that

$$\mathbf{v}_\alpha \cdot \frac{\partial n_\alpha}{\partial \mathbf{r}} = v_x \frac{x}{r}\frac{\partial n_\alpha}{\partial r} + v_y \frac{y}{r}\frac{\partial n_\alpha}{\partial r} + v_z \frac{z}{r}\frac{\partial n_\alpha}{\partial r}. \tag{17}$$

Substituting (17) to (14) we obtain

$$\boldsymbol{\Gamma}_{\alpha,D} = -\frac{1}{n_{\alpha 0}\nu_\alpha}\int d\mathbf{v}_\alpha \left[\left(v_x^2 \frac{x}{r}\frac{\partial n_\alpha}{\partial r}\right)\boldsymbol{i} + \left(v_y^2 \frac{y}{r}\frac{\partial n_\alpha}{\partial r}\right)\boldsymbol{j} + \left(v_z^2 \frac{z}{r}\frac{\partial n_\alpha}{\partial r}\right)\boldsymbol{k}\right] f_{\kappa,\alpha}. \tag{18}$$

Furthermore, we find that

$$\int v_x^2 f_{\kappa,\alpha}\frac{x}{r}\frac{\partial n_\alpha}{\partial r}d\mathbf{v}_\alpha = \int v_y^2 f_{\kappa,\alpha}\frac{y}{r}\frac{\partial n_\alpha}{\partial r}d\mathbf{v}_\alpha = \int v_z^2 f_{\kappa,\alpha}\frac{z}{r}\frac{\partial n_\alpha}{\partial r}d\mathbf{v}_\alpha$$

$$= \frac{4\pi}{3}\nabla n_\alpha \int v_\alpha^2 f_{\kappa,\alpha} v_\alpha^2 dv_\alpha. \tag{19}$$

And then,

$$\boldsymbol{\Gamma}_{\alpha,D} = -\frac{4\pi}{3n_{\alpha 0}\nu_\alpha}\nabla n_\alpha \int v_\alpha^4 f_{\kappa,\alpha} dv_\alpha. \tag{20}$$

Or it can be written as

$$\boldsymbol{\Gamma}_{\alpha,D} = -\frac{4\pi B_{\kappa,\alpha}}{3\nu_\alpha}\frac{\partial n_\alpha}{\partial r}\int v_\alpha^4 \left(1 + A_{\kappa,\alpha}{v_\alpha}^2\right)^{-(\kappa_\alpha+1)} dv_\alpha. \tag{21}$$

After calculating the integral in Eq.(21), we obtain that

$$\boldsymbol{\Gamma}_{\alpha,D} = -\frac{2\pi^2 \left(2\kappa_\alpha - 5\right)!!}{\nu_\alpha A_{\kappa,\alpha}^{5/2}\left(2\kappa_\alpha\right)!!} B_{\kappa,\alpha}\nabla n_\alpha. \tag{22}$$

As compared with the Fick's diffusion law,

$$\boldsymbol{\Gamma}_{\alpha,D} = -D_\alpha \nabla n_\alpha, \tag{23}$$

we find that the diffusion coefficient $D_{\kappa,\alpha}$ of $\alpha$th particle (i.e. electrons and ions) in the weakly ionized plasma with $\kappa$-distributions is expressed by

$$\begin{aligned} D_{\kappa,\alpha} &= \frac{1}{\nu_\alpha}\frac{2\pi^2 B_{\kappa,\alpha}}{A_{\kappa,\alpha}^{5/2}}\frac{\left(2\kappa_\alpha - 5\right)!!}{\left(2\kappa_\alpha\right)!!} \\ &= \frac{2\sqrt{\pi}k_B T_\alpha \left(2\kappa_\alpha - 3\right)}{m_\alpha \nu_\alpha}\frac{\Gamma\left(\kappa_\alpha + 1\right)}{\Gamma\left(\kappa_\alpha - 1/2\right)}\frac{\left(2\kappa_\alpha - 5\right)!!}{\left(2\kappa_\alpha\right)!!}. \end{aligned} \tag{24}$$

Namely, for the electrons, the diffusion coefficient is

$$D_{\kappa,e}=\frac{2\sqrt{\pi}k_BT_e\left(2\kappa_e-3\right)}{m_e\nu_e}\frac{\Gamma\left(\kappa_e+1\right)}{\Gamma\left(\kappa_e-1/2\right)}\frac{\left(2\kappa_e-5\right)!!}{\left(2\kappa_e\right)!!},\tag{25}$$

and for the ions, the diffusion coefficient is

$$D_{\kappa,i}=\frac{2\sqrt{\pi}k_BT_i\left(2\kappa_i-3\right)}{m_i\nu_i}\frac{\Gamma\left(\kappa_i+1\right)}{\Gamma\left(\kappa_i-1/2\right)}\frac{\left(2\kappa_i-5\right)!!}{\left(2\kappa_i\right)!!}\tag{26}$$

It is clear that the new diffusion coefficients depend strongly on the $\kappa$-parameter, and when we take $\kappa_\alpha\to\infty$, they can reduce to the standard form for the plasma following a Maxwellian distribution,[16]

$$D_\alpha=\frac{k_BT_\alpha}{m_\alpha\nu_\alpha}.\tag{27}$$

## IV. The mobility of charged particles

In Eq.(13) we have know that the second term is due to electric field driven particle migration flow. We now study the mobility of electrons and ions in the weakly ionized plasma with $\kappa$-distributions.

The charged particles driven by the electric field through a medium of target particles will collide with them, and will move in the direction parallel to the electric field. Such migration flow can be described by the transport coefficient called as mobility. In the weakly ionized plasma without magnetic field, when the electrons and ions are passing through the medium filled with the neutral particles and few charged particles, they will collide with each other, and also the charged particles will be driven by the electric field force. According to Eq.(13), if the migration flow of charged particles is denoted by $\Gamma_{\alpha,\mu}$, then we have that

$$\begin{aligned}\mathbf{\Gamma}_{\alpha,\mu}&=-\int\frac{\mathbf{a}_\alpha}{\nu_\alpha}\cdot\frac{\partial}{\partial\mathbf{v}_\alpha}\left(\frac{n_\alpha}{n_{\alpha0}}f_{\kappa,\alpha}\right)\mathbf{v}_\alpha d\mathbf{v}_\alpha\\&=-\frac{n_\alpha Z_\alpha e\mathbf{E}}{3n_{\alpha0}m_\alpha\nu_\alpha}\int_0^\infty 4\pi v_\alpha^3\frac{\partial f_{\kappa,\alpha}}{\partial v_\alpha}dv_\alpha\\&=\frac{4\pi n_\alpha Z_\alpha e\mathbf{E}}{3n_{\alpha0}m_\alpha\nu_\alpha}\left[-v_\alpha^3 f_{\kappa,\alpha}\,|_0^\infty+\int_0^\infty f_{\kappa,\alpha}d\left(v_\alpha^3\right)\right].\end{aligned}\tag{28}$$

The first term in the brackets equals to zero, because in the limit of the speed $v_\alpha\to\infty$, $f_{\kappa,\alpha}\to 0$ is faster than $v^3\to\infty$. Thus Eq.(28) becomes

$$\mathbf{\Gamma}_{\alpha,\mu}=\frac{4\pi n_\alpha Z_\alpha e\mathbf{E}}{n_{\alpha0}m_\alpha\nu_\alpha}\int_0^\infty v_\alpha^2 f_{\kappa,\alpha}dv_\alpha\tag{29}$$

Submiting the $\kappa$-distribution function (7) into Eq.(29), and calculating the integral in this equation, we obtain that

$$\mathbf{\Gamma}_{\alpha,\mu} = \frac{2\pi^2 n_\alpha Z_\alpha e \mathbf{E} B_{\kappa,\alpha} \left(2\kappa_\alpha - 3\right)!!}{m_\alpha \nu_\alpha A_{\kappa,\alpha}^{3/2} \left(2\kappa_\alpha\right)!!}. \tag{30}$$

According to the definition of mobility, the formula is

$$\mathbf{\Gamma}_{\alpha,\mu} = n_\alpha \mu_\alpha \mathbf{E}, \tag{31}$$

and then we can derive the expressions of mobility coefficient $\mu_{\kappa,\alpha}$ of charged particles (i.e. electrons and ions) in the weakly ionized plasma with $\kappa$-distribution,

$$\begin{aligned}\mu_{\kappa,\alpha} &= \frac{2\pi^2 z_\alpha e B_{\kappa,\alpha} \left(2\kappa_\alpha - 3\right)!!}{m_\alpha \nu_\alpha A_{\kappa,\alpha}^{3/2} \left(2\kappa_\alpha\right)!!} \\ &= \frac{2\sqrt{\pi} Z_\alpha e \left(2\kappa_\alpha - 3\right)!! \Gamma\left(\kappa_\alpha + 1\right)}{m_\alpha \nu_\alpha \left(2\kappa_\alpha\right)!! \Gamma\left(\kappa_\alpha - 1/2\right)}.\end{aligned} \tag{32}$$

We also see that the new mobility coefficient depends strongly on the $\kappa$-parameter, and in the limit of $\kappa_\alpha \to \infty$, it can be reduced to the classical form in the weakly ionized plasma following a Maxwellian distribution, i.e.,[16]

$$\mu_\alpha = \frac{Z_\alpha e}{m_\alpha \nu_\alpha}. \tag{33}$$

**V. Conclusion and discussion**

In conclusion, we have studied the diffusion of charged particles in the weakly ionized plasma with the power-law $\kappa$-distributions and without magnetic field. The electrons and ions have different $\kappa$-parameter. We have derived the expressions of the diffusion coefficients of electrons and ions, respectively, in the weakly ionized plasma with the $\kappa$-distributions. They are Eq.(25) and Eq.(26). We find that the new diffusion coefficients depend strongly on the $\kappa$-parameter of the power-law distributed plasma.

We have also derived the expressions of the mobility coefficients of electrons and ions in the weakly ionized plasma with the $\kappa$-distributions. They are given by Eq.(32), which also depends strongly on the $\kappa$-parameter of the power-law distributed plasma.

When we take $\kappa \to \infty$, all the equations can reduce to the forms in the weakly ionized plasma with a Maxwellian distribution.

In addition, it is important for us to understand the physical meaning of the $\kappa$-parameter in the plasmas with power-law distributions. Now one knows that in the plasmas the $\kappa$-parameter can be determined generally by the following equation,[10]

$$(\kappa_\alpha - \frac{3}{2}) k_B \nabla T_\alpha = e\,[-\nabla \varphi_c + c^{-1} \boldsymbol{u} \times \boldsymbol{B}] - m_\alpha (\omega^2 \boldsymbol{R} + 2\boldsymbol{u} \times \boldsymbol{\omega}), \tag{34}$$

where $\varphi_c$ is the Coulomb potential, $c$ is the light speed, $\boldsymbol{B}$ is the magnetic induction intensity. The rotation effect contains two terms, $m_\alpha \omega^2 \boldsymbol{R} + 2m_\alpha \boldsymbol{u} \times \boldsymbol{\omega}$. The first term $m_\alpha \omega^2 \boldsymbol{R}$ is the contribution due to the inertial centrifugal force. When the angle velocity $\boldsymbol{\omega}$ varies from the equator to poles, the differential rotation exists depending on the angle velocity $\boldsymbol{\omega}$ and the vertical distance $\boldsymbol{R}$ between the particle and the rotation axis. The second term $2m_\alpha \boldsymbol{u} \times \boldsymbol{\omega}$ is the contribution due to the Coriolis force, depending on the angle velocity $\boldsymbol{\omega}$ and the convective motion velocity $\boldsymbol{u}$ of the plasma fluid. It is shown that the $\kappa$-parameter is related not only to the temperature gradient, but also to the electromagnetic fields and the rotation in the plasma. For the $\kappa$-parameter of the plasmas in the case without the magnetic field and without the rotation, we just need to take $\boldsymbol{B}$=0 and $\omega$=0 in Eq.(34).

**Acknowledgment**

This work was supported by the National Natural Science Foundation of China under Grant No. 11775156.

**Reference**

[1] V. M. Vasyliunas, J.Gerophys. Res. **73**, 2839 (1968).
[2] J. D. Mihalov, J. H. Wolfe, and L. A. Frank, J. Geophys. Res. **81**, 3412 (1976).
[3] S. R. Cranmer, Astrophys. J. **508**, 925 (1998).
[4] M. P. Leubner, Astrophys. J. **604**, 469 (2004); Astrophys. Space Sci. **282**, 573 (2002).
[5] L. Y. Liu and J. L. Du, Physica A **387**, 4821 (2008) ; J. L. Du, Phys. Lett. A **329**, 262 (2004).
[6] Z. P. Liu and J. L. Du, Phys. Plasmas **16**, 123707 (2009).
[7] J. L. Du, Phys. Plasmas **20**, 092901 (2013).
[8] G. Livadiotis, J. Geophys. Res. **120**, 880 (2015).
[9] G. Livadiotis, J. Geophys. Res. **120**, 1607 (2015).
[10] H. N. Yu and J. L. Du, EPL **116** (2016) 60005.
[11] M. Bacha, L. A. Gougam and M. Tribeche, Physica A **466**, 199 (2017).
[12] C. Tsallis, see the website at http://tsallis.cat.cbpf.br/biblio.htm
[13] K. Rohlena and H. R. Skullerud, Phys. Rev. E **51**, 6028 (1995).
[14] N. A. Krall and A. W. Trivelpiece, *Principles of Plasma Physics* (McGraw-Hill Book Company, 1973).
[15] J. L. Xu and S. X. Jin, *Plasma Physics,* the front page.(Atomic Energy Press, Beijing, 1981).
[16] N.A.Krall and A.W.Trivelpie, *The Principle of Plasma Physics*, the third edition. (Atomic Energy Press, Beijing, 1983).

# Erratum: "The diffusion of charged particles in the weakly ionized plasma with power-law kappa-distributions" [Phys. Plasmas 24, 102305 (2017)]

Lan Wang and Jiulin Du
*Department of Physics, School of Science, Tianjin University, Tianjin 300072, China*

In the published version of the paper,[1] errors are found in the calculations of Eq.(22) and Eq.(30), respectively. The diffusion flow,[1] Eq.(22),

$$\mathbf{\Gamma}_{\alpha,D} = -\frac{2\pi^2 (2\kappa_\alpha - 5)!!}{\nu_\alpha A_{\kappa,\alpha}^{5/2} (2\kappa_\alpha)!!} B_{\kappa,\alpha} \nabla n_\alpha \ ,$$

should be corrected as

$$\begin{aligned}\mathbf{\Gamma}_{\alpha,D} &= -\frac{2\pi B_{\kappa,\alpha}}{3\nu_\alpha A_{\kappa,\alpha}^{5/2}} \mathrm{B}\left(\frac{5}{2}, \kappa - \frac{3}{2}\right) \nabla n_\alpha \\ &= -\frac{2\pi B_{\kappa,\alpha}}{3\nu_\alpha A_{\kappa,\alpha}^{5/2}} \frac{\Gamma\left(\frac{5}{2}\right)\Gamma\left(\kappa - \frac{3}{2}\right)}{\Gamma(\kappa+1)} \nabla n_\alpha \\ &= -\frac{k_B T_\alpha}{m_\alpha \nu_\alpha} \nabla n_\alpha , \end{aligned} \tag{E1}$$

where B( a, b) is the Beta function. Correspondingly, the diffusion coefficients Eqs.(24) -(26) should be corrected as

$$D_{\kappa,\alpha} = \frac{k_B T_\alpha}{m_\alpha \nu_\alpha} = D_{\infty,\alpha} \ . \tag{E2}$$

In the paper,[1] the migration flow, Eq.(30),

$$\mathbf{\Gamma}_{\alpha,\mu} = \frac{2\pi^2 n_\alpha Z_\alpha e \mathbf{E} B_{\kappa,\alpha} (2\kappa_\alpha - 3)!!}{m_\alpha \nu_\alpha A_{\kappa,\alpha}^{3/2} (2\kappa_\alpha)!!} \ ,$$

should be corrected as

$$\mathbf{\Gamma}_{\alpha,\mu} = \frac{n_\alpha Z_\alpha e}{m_\alpha \nu_\alpha} \mathbf{E} \ . \tag{E3}$$

Correspondingly, the mobility coefficient,[1] Eq.(32), should be corrected as

$$\mu_{\kappa,\alpha} = \frac{Z_\alpha e}{m_\alpha \nu_\alpha} = \mu_{\infty,\alpha} \ . \tag{E4}$$

In (E2) and (E4), $D_{\infty,\alpha}$ and $\mu_{\infty,\alpha}$ are respectively the coefficients for the case $\kappa \to \infty$.

In conclusion, in the weakly ionized plasma with the power-law $\kappa$-distributions and without magnetic field, if the collision frequencies $\nu_\alpha$ were assumed to be constant, the diffusion and mobility coefficient of charged particles would not be dependent on the $\kappa$-parameter, so they would be both the same as those in the plasma with a Maxwellian distribution.

We acknowledge Ms. Xiaoxue Ji for the calculation in Eq.(E1). This work is supported by National Natural Science Foundation of China under Grant No.11775156.

[1] L.Wang and J. Du, Phys. Plasmas 24, 102305 (2017).